\documentclass{pj}
\usepackage{upgreek}
\usepackage{combelow}
\usepackage{comment}
\usepackage{cite}
\usepackage{tikz,graphicx}
\usepackage{amsmath}
\usepackage{color}
\usepackage[mathscr]{eucal}
\usepackage{amssymb}

\def\XXint#1#2#3{{\setbox0=\hbox{$#1{#2#3}{\int}$}
     \vcenter{\hbox{$#2#3$}}\kern-.5\wd0}}

\newcommand{\be}[1]{\begin{equation} \label{#1} }
\newcommand{\bea}[1]{\begin{eqnarray} \label{#1} }

\newcommand{\bfi}{\begin{figure}}
\newcommand{\efi}{\end{figure}} 
\newcommand{\ee}{\end{equation}}
\newcommand{\eea}{\end{eqnarray}}
\newcommand{\bib}{\bibitem}
\newcommand{\lbl}{\label}

\newcommand{\vc}{{\bf c}}

\newcommand{\vv}{{\bf v}}

\newcommand{\vH}{{\bf H}}
\newcommand{\vE}{{\bf E}}
\newcommand{\vM}{{\bf M}}
\newcommand{\vP}{{\bf P}}

\newcommand{\vA}{{\bf A}}
\newcommand{\vB}{{\bf B}}

\newcommand{\vD}{{\bf D}}
\newcommand{\vJ}{{\bf J}}

\newcommand{\vnh}{{\bf \hat{n}}}

\newcommand{\eps}{\epsilon}

\begin{document}
\footnote{}  \footnote{} \footnote{}
\title[]{\mbox{}\\[-20mm]Maxwell's definition of electric polarization as displacement}
\author{Arthur~D.~Yaghjian
}
\runningauthor{}
\par
\begin{abstract}
After reaffirming that the macroscopic dipolar electromagnetic equations, which today are commonly referred to as Maxwell's equations, are found in Maxwell's Treatise, we explain from his Treatise that Maxwell defined his displacement vector $\vD$ as the electric polarization and did not introduce in his Treatise or papers the concept of electric polarization $\vP$ or the associated electric-polarization volume and surface charge densities, $-\nabla\cdot\vP$ and $\vnh\cdot\vP$, respectively.  With this realization, we show that Maxwell's discussion of surface charge density between volume elements of dielectrics and between dielectrics and conductors becomes understandable and valid within the context of his definition of electric polarization as displacement $\vD$.  Apparently, this identification of $\vD$ with electric polarization in Maxwell's work has not been previously pointed out or documented except very briefly in \cite{Yaghjian-Reflection}. 
\end{abstract}
%
%
\section{Introduction}
One of the important features of Maxwell's Treatise \cite{Maxwell}, as mentioned in the paper \cite{Yaghjian-Reflection}, is Maxwell's introduction of the electric displacement vector $\vD$ without ever defining an electric polarization vector $\vP$ or assuming that the change in electric displacement in dielectrics from that in free space is caused by electric dipoles.  In stark contrast, Maxwell thoroughly analyzes magnetization $\vM$ (magnetic polarization) in terms of elementary magnetic dipoles produced by hypothetical magnetic charge separation and introduces the constitutive relation $\vB = \mu_0(\vH+\vM)$ with $\vH$ the primary magnetic field produced by the hypothetical magnetic charge and $\mu_0$ the permeability of free space. (In the present paper, all equations are written using modern SI units.)   As explained in \cite{Yaghjian-Reflection}, although Maxwell assumes that there is some mechanism for charge separation that produces the displacement vector, he does not assume that this induced charge separation in a dielectric is caused by electric dipoles within the dielectric that give rise to a polarization vector $\vP$ or polarization volume and surface charge densities, $-\nabla\cdot\vP$ and $\vnh\cdot\vP$, respectively. He limits his description of dielectrics to the linear constitutive relation $\vD = \eps\vE$ \cite[Art. 68]{Maxwell} with $\nabla\cdot\vD=\rho_e$ \cite[Arts. 83a and 612]{Maxwell} (equations that he infers from Faraday's experimental results) but without suggesting the possibility of the more general constitutive relation $\vD = \eps_0\vE+ \vP$, where $\vE$ is the electric field, $\vD$ is the displacement, $\eps$ is the permittivity (which Maxwell allowed to be a tensor) of the dielectric, $\eps_0$ is the free-space permittivity, and $\rho_e$ is the electric charge density.
\par
Probably the two main reasons that Maxwell does not introduce an electric polarization vector $\vP$ analogous to his magnetic polarization vector $\vM$ is 1) relatively very little experimentation had been done with electrets (compared to the vast amount of experimentation with magnets) to strongly indicate the dipole nature of electric polarization and 2) Maxwell did not know beforehand the form of Ampere's law with the $\eps_0\partial\vE/\partial t$ term when he inserted the $\partial \vD/\partial t$ term to make Ampere's static law ($\nabla\times\vH = \vJ$, where $\vJ$ is the current density) compatible with the continuity equation ($\nabla\cdot\vJ = -\partial \rho_e/\partial t$) and the divergence equation ($\nabla\cdot\vD = \rho_e$).  Incidentally, it was Maxwell, not Ampere, who first deduced $\nabla\times\vH = \vJ$ in his 1856 (part 1 was read December 1855) paper ``On Faraday's Lines of Force'' \cite{Maxwell-1856}.
\par
It is well-known that Maxwell referred to $\partial \vD/\partial t$ as displacement current and $\vJ+\partial \vD/\partial t$, rather than $\vJ+\partial \vP/\partial t$, as the total current.  However, although mentioned in Section 4.1 of \cite{Yaghjian-Reflection}, it has not been documented with enough detail that Maxwell used the term ``electric polarization'' synonymously with the term ``electric displacement'' or just ``displacement.''   In other words, when Maxwell talks about the electric polarization, he is referring to the vector $\vD$ and not some imaginary vector $\vP$ that is never defined in his Treatise or papers.  When one realizes this important definition, interpretation and word usage of Maxwell, several of his statements about electric polarization that seem confusing (if one thinks of $\vP$ as electric polarization) become fully understandable when one considers (as Maxwell did) that $\vD$ is the measure of electric polarization, whether in a dielectric or the ether, the ``luminiferous medium'' that Maxwell assumed filled a vacuum \cite[Arts. 782 and 866]{Maxwell}.\footnote{Although a few references have noted that Maxwell does not introduce a vector $\vP$, I have not been able to find a single reference that realizes that when Maxwell uses the term or refers to  ``electric polarization'' he means the vector $\vD$ and does not have in mind our present-day concept of electric polarization $\vP$, which was introduced many years after Maxwell died.  This misunderstanding of Maxwell continues to be a source of confusion for both scientists and historians of science especially when they try to interpret what Maxwell says about the charges at the surfaces of dielectrics and at the interfaces of dielectrics and conductors \cite[Ch. 3]{Buchwald}, \cite{HW}; see Section \ref{SC} below.}  The documentation and clarification of this important feature of Maxwell's Treatise (and papers) is the primary objective of the present paper.  As a secondary objective, we review in the next section the principal equations contained in Maxwell's Treatise.
\section{Maxwell, Founder of Maxwell's Equations}
The most common present-day expression of the Maxwellian macroscopic electromagnetic field equations in dipolar media is exemplified in Chapter 1 of Stratton's book \cite{Stratton} on \textit{Electromagnetic Theory}, namely
\be{1}
\nabla\times\vE = -\frac{\partial \vB}{\partial t}
\ee
\be{2}
\nabla\times\vH = \vJ +\frac{\partial \vD}{\partial t}
\ee
\be{3}
\nabla\cdot\vB = 0
\ee
\be{4}
\nabla\cdot\vD = \rho_e
\ee
with constitutive relations
\be{5}
\vD = \eps_0\vE + \vP
\ee
\be{6}
\vB = \mu_0(\vH + \vM)
\ee
and for isotropic, linear, stationary materials
\be{7}
\vD = \eps\vE
\ee
\mbox{}\\[-10mm]
\be{8}
\vB = \mu\vH
\ee
\be{9}
\vJ = \sigma\vE.
\ee
\mbox{}\\[-6mm]
Each one of the vector-form equations in (\ref{1})--(\ref{9}) are found in Art. 619 of Maxwell's Treatise (after changing to SI units and Maxwell's German-script vectors to modern boldface vectors) with the exceptions of (\ref{1}), (\ref{3}) and (\ref{5}).
\par
As explained in the Introduction, Maxwell did not introduce the vector $\vP$ and thus did not introduce equation (\ref{5}).  Nevertheless, his polarization vector $\vD$ in (\ref{2}) and (\ref{7}) was quite general in that Maxwell assumed that the permittivity, as well as the permeability and conductivity, could be tensors and, thus, the material could be anisotropic \cite[Arts. 101e, 298, 428, 608, 609, 794]{Maxwell}.
\par
The three scalar equations for the rectangular components of the mathematical rendition of ``Faraday's law'' in (\ref{1}) can be found as equation (54) in Maxwell's 1861 \textit{Philosophical Magazine} paper, ``On Physical Lines of Force'' \cite{Maxwell1861}, \cite[Ch. 4]{HW}.  In Maxwell's Treatise, equation (\ref{1}) appears in its integral form \cite[Arts. 579 and 595]{Maxwell}
\begin{subequations}
\lbl{10}
\be{10a}
E = -\frac{dp}{dt}
\ee
with $E$ and $p$ defined in terms of integrals of the vectors $\vE$ and $\vB$ in Arts. 598 and 591 of the Treatise
\be{11a}
E = \oint\limits_C \vE\cdot d\vc,\;\;\;\; p = \int\limits_S \vB\cdot\vnh\, dS.
\ee
\end{subequations}
(Maxwell worked extensively with (\ref{10}) in Art. 598 where he shows that the force $\vv\times\vB$ must be added to $\vE$ if the curve $C$ is moving with velocity $\vv$ and this implies that the force exerted on a hypothetical unit electric charge moving with $C$ is $\vE+\vv\times\vB$.  Thus, Maxwell derived the ``Lorentz force'' from his generalized equation for Faraday's law \cite{Yaghjian-Reflection}, \cite{Yaghjian-arXiv}.)  
\par
Maxwell wrote equation (\ref{3}) several times in his Treatise as well as in his papers but always in the scalar form\footnote{Maxwell uses the derivative symbol ``$d$'' rather than the partial derivative symbol ``$\partial$'', which was not in wide use during his lifetime.  Nonetheless, he specified that, for example, $da(x,y,z)/dx$ means the derivative of $a(x,y,z)$ with respect to $x$ holding $y$ and $z$ fixed.}
\be{3'}
\frac{\partial B_x}{\partial x} + \frac{\partial B_y}{\partial y} + \frac{\partial B_z}{\partial z} = 0.
\ee
\par
It should be noted that Maxwell did not obtain his macroscopic equations by averaging the fields of discrete charges or dipoles but by assuming that each infinitesimal volume element $dV$ of material (or ether) has the same electromagnetic properties as a continuum.  For example, for magnetization, he assumes the dipole moment of the  volume element is $\vM dV$.  He derives the static equations first and then generalizes to the time varying equations.  Exactly how he does this, distinguishing between mathematically defined fields and measurable fields, is explained in \cite{Yaghjian-Reflection}.  
\par
To my knowledge, none of the equations (\ref{1})--(\ref{9}) can be found (in scalar or vector differential or integral form) in any publications previous to Maxwell's papers and Treatise, except for (\ref{3}), which W. Thomson determined for magnetostatic fields in his 1851 paper \cite{Thomson}.  Moreover, without the papers and Treatise of Maxwell, it is highly unlikely that other scientists of his time would have inferred in their lifetimes these equations from Faraday's \textit{Experimental Researches in Electricity} \cite{Faraday}, which doesn't contain a single equation.   It took the singular genius of Maxwell --- his phenomenal powers of thought and physical intuition coupled with extraordinary mathematical skills --- to deduce these remarkable equations.  It seems to me a travesty to refer to these equations other than ``Maxwell's equations.''  Nonetheless, it would certainly be unfair to Faraday not to refer to the physics that (\ref{1}) and (\ref{10}) represent mathematically as Faraday's law.\footnote{A ``law'' of physics is used here to mean the verbal statement of the relationship between measured quantities during  repeatable physical experiments.  An ``equation'' of physics is used to mean the formulation in mathematical symbols of the measured quantities and their relationship.  For example, Faraday summarized his ``law'' of electromagnetic induction as ``A piece of metal or conducting matter which moves across lines of magnetic force has, or tends to have a current of electricity produced in it'' \cite[paragraph 3087, p. 335]{Faraday}.  The corresponding Maxwell ``equation'' is $\oint_C \vE\cdot d\vc = -\int_S \partial\vB/\partial t\cdot\vnh\, dS$.} In the Preface and throughout his Treatise, Maxwell gives full credit for the experiments and ideas underlying most of his equations to Faraday.  Indeed, Maxwell could not have determined his equations without the previous work of Faraday, Oersted, Ampere, Weber, Thomson and others.
\par
Although Maxwell is generally considered the most important physical scientist between Newton and Einstein,  
 unfortunately too many commentaries on Maxwell's work have claimed that Maxwell didn't express his equations in the form (\ref{1})--(\ref{9}) that we know them today.  Probably the two main reasons for that misunderstanding are  1)  Maxwell's definitive 1865 \textit{Philosophical Transactions} paper \cite{Maxwell1865} contained (in Maxwell's words) ``twenty variable [scalar] quantities'' and ``between these twenty [scalar] quantities we have found twenty [scalar] equations'' and 2) in Art. 619 of Maxwell's Treatise, where he summarizes his general theory for the dynamic electromagnetic field with twelve  vector-form  equations, he uses instead of (\ref{1}) and (\ref{3}) the equations
\be{12}
\vB = \nabla \times \vA
\ee
and
\be{13}
\vE = -\frac{\partial \vA}{\partial t} -\nabla\psi_e
\ee
where $\vA$ and $\psi_e$ are the vector potential and the scalar electric potential.  Of course, taking the divergence of (\ref{12}) yields (\ref{3}) and taking the curl of (\ref{13}) with substitution from (\ref{12}) yields (\ref{1}).  And, as emphasized above and detailed in \cite{Yaghjian-Reflection}, the equation (\ref{1}) (in integral form) and the equation (\ref{3}) (in scalar form) are also contained and worked with extensively in other Articles of Maxwell's Treatise leading up to Art. 619.   The other equations, (\ref{2}), (\ref{4}) and (\ref{6})--(\ref{9}), are found in Art. 619 along with three additional equations that Maxwell includes, namely his ``equation of mechanical force'' that gives the force density on current carrying conductors, polarized dielectrics, and bodies with static charge or static magnetization (see equation (\ref{mf})); his definition of the ``total current'' as the sum of the conduction current and the time rate of change of electric displacement ($\vJ_T = \vJ + \partial \vD/\partial t$); and the equation $\vH = -\nabla \psi_m$ when the magnetic field can be derived from a scalar potential. 
\section{Maxwell's Electric Polarization Vector is Displacement $\vD$}\lbl{MEPV}
As we mentioned in the Introduction, Maxwell's Treatise (and papers) have no concept of what today is call the electric polarization $\vP$, which was introduced at the turn of the century by Larmor \cite{Larmor}, Leatham \cite{Leathem}, and Lorentz \cite{Lorentz}.  Thus there is also no concept of electric-polarization volume and surface charge densities, $-\nabla\cdot\vP$ and $\vnh\cdot\vP$, in Maxwell's work, but only what we refer to today as electric charge density $\rho_e$ satisfying the continuity equation, $\nabla\cdot\vJ = -\partial \rho_e/\partial t$.  In Maxwell's Treatise, this electric charge density \cite[Art. 31]{Maxwell}, which he also calls ``electrification,'' or ``free electricity,'' or just ``electricity,'' is a fluid substance (continuum) \cite[Art. 36]{Maxwell}; and electric conduction current $\vJ$ is the ``transference of electrification'' \cite[Art. 231]{Maxwell}. 
\par
Throughout vol. I of his Treatise, Maxwell defines and refers to electric polarization as the displacement $\vD$. (In vol. II, Maxwell does not discuss electric polarization per se --- just displacement.) He begins his discussion of electric polarization in Art. 59 of his Treatise where he says, ``It is better, however, in considering the theory of dielectrics from the most general point of view, to distinguish between the electromotive intensity [$\vE$] at any point and the electric polarization  of the medium at that point, since these directed quantities, though related to one another, are not, in some solid substances, in the same direction. The most general expression for the electric energy of the medium per unit of volume is half the product of the electromotive intensity [$\vE$] and the electric polarization [$\vD$] multiplied by the cosine of the angle between their directions [$\vE\cdot\vD/2$ -- see Art. 111]. In all fluid dielectrics the electromotive intensity and the electric polarization are in the same direction and in a constant ratio [$\vD=\eps\vE$ -- see Arts. 68 and 111].''  
\par
In Art. 60, Maxwell declares again that electric polarization is synonymous with electrical displacement: ``The electric polarization of an elementary portion of a dielectric is a forced state into which the medium is thrown by the action of electromotive force, and which disappears when that force is removed. We may conceive it to consist in what we may call an electrical displacement [$\vD$], produced by the electromotive intensity [$\vE$].  When the electromotive force acts on a conducting medium it produces a current through it, but if the medium is a nonconductor or dielectric, the current cannot flow through the medium, but the electricity is displaced within the medium in the direction
of the electromotive intensity, the extent of this displacement depending on the magnitude of the electromotive intensity [$\vD=\eps\vE$], so that if the electromotive intensity increases or diminishes, the electric displacement increases and diminishes in the same ratio.  \textit{The amount of the displacement is measured by the quantity of electricity which crosses a unit of area, while the displacement increases from zero to its actual amount. This, therefore, is the measure of the electric polarization.}''
\par
As a consequence of Maxwell's definition of the measure of electric polarization as the vector $\vD$, the electric polarization is zero wherever $\vD$ is zero (not where the present-day electric polarization vector $\vP$ is zero).  This is further confirmed by the words of Maxwell in Art. 62, ``That the energy of electrification resides in the dielectric medium, whether that medium be solid, liquid, or gaseous, dense or rare, \textit{or even what is called a vacuum}, provided it be still capable of transmitting electrical action.  \textit{That the energy in any part of the medium is stored up in the form of a state of constraint called electric polarization, the amount of which} [$\vD$] \textit{depends on the resultant electromotive intensity} [$\vE$] \textit{at the place} [$\vD=\eps\vE$].''  In other words, according to Maxwell's definition of electric polarization, there exists electric polarization even in the vacuum/ether if $\vD$ is not equal to zero.  Therefore, whenever Maxwell says that there is zero electric polarization in a solid dielectric or vacuum/ether, he means that $\vD$ is zero in that region.  Much of what Maxwell says about electric polarization in his Treatise becomes understandable only within the realization that Maxwell uses the term ``electric polarization'' synonymously with ``electric displacement'' or just ``displacement.''  Maxwell's expression
\be{mf}
\left(\vJ+\frac{\partial\vD}{\partial t}\right)\times\vB
\ee
in Art. 619 for the ``mechanical force'' exerted by the magnetic induction $\vB$ on a current carrying conductor or polarized dielectric is consistent with his assumption that $\vD$ is the measure of electric polarization even in the vacuum/ether where $\vD=\eps_0\vE$. (Today we would properly express this mechanical force as $(\vJ+\partial\vP/\partial t)\times\vB$ with $\vP = (\eps -\eps_0)\vE$ such that the electric polarization force is zero in a vacuum where $\eps = \eps_0$ \cite[Sec. 2.1.10]{H&Y}.) 
\section{\lbl{SC}Maxwell's Surface Charge on Dielectrics produced by Electric Polarization}
With Maxwell's definition of electric polarization as the displacement $\vD$, along with his divergence equation $\nabla\cdot\vD = \rho_e$ \cite[Arts. 83a or 612]{Maxwell} and his normal-displacement boundary condition $D_{n2}+D_{n1}= \sigma_e$ (with $\vnh_{1,2}$ defined into the media on either side of the surface) \cite[Arts. 83a or 613]{Maxwell}, where $\sigma_e$ is the surface charge density, his discussions of the surface charge associated with electric polarization in dielectrics --- discussions that have been a source of confusion to the present day  \cite[Ch. 3]{Buchwald}, \cite{HW} --- become understandable.  
\par
\begin{figure}[htbp]
  \centering
  \includegraphics[width=80mm]{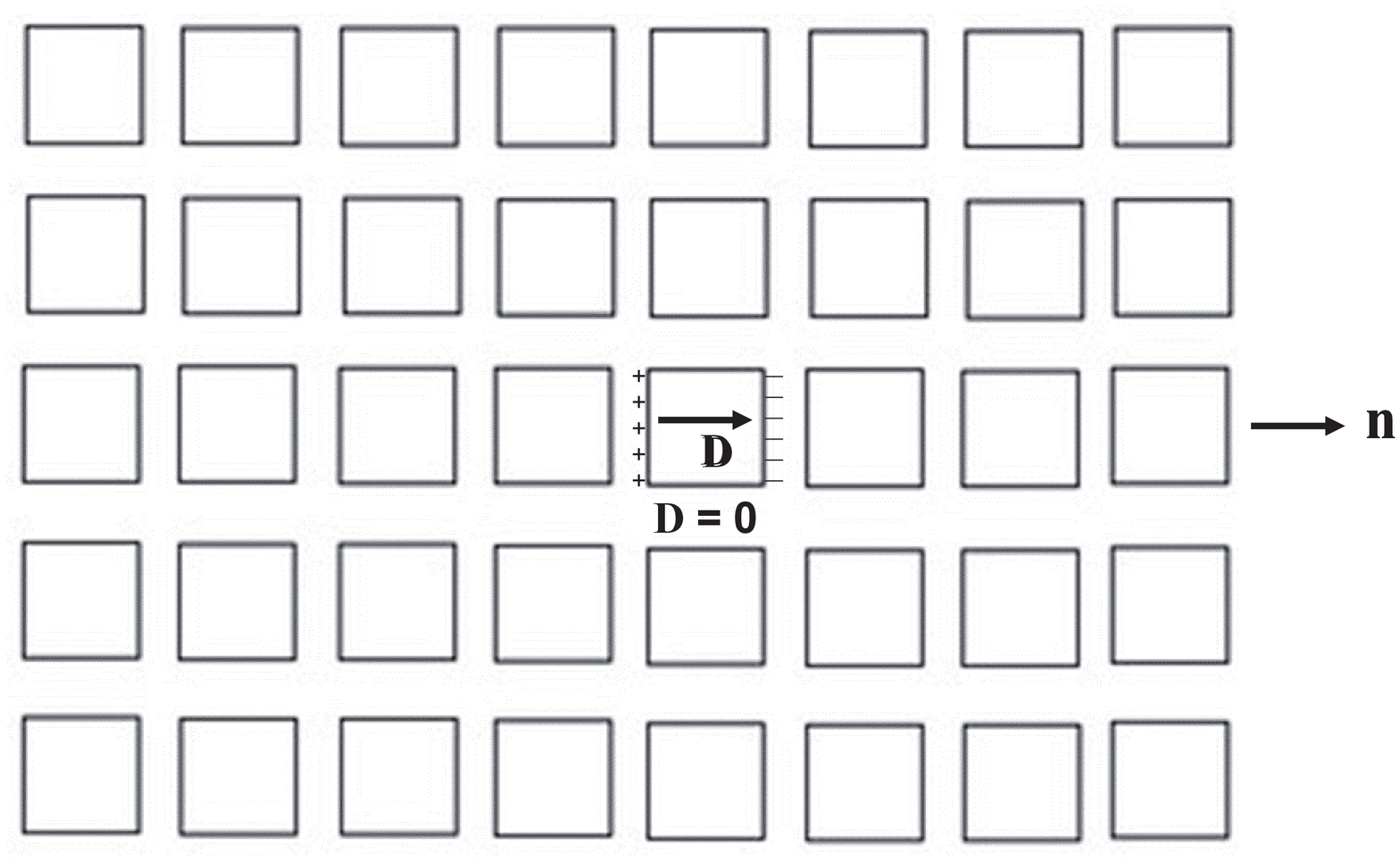}\\[-3mm]
  \caption{\small Cubical volume elements of a dielectric having displacement $\vD$ (Maxwell electric polarization) separated by thin shells of zero displacement (zero Maxwell electric polarization).  The boundary condition on normal $\vD$ requires that $\sigma_e = -D_n$ on the positive (right) side of the volume elements and that $\sigma_e = +D_n$ on the negative (left) side of the volume elements.}
  \label{figVolumeElements}
\end{figure}

For example, consider Maxwell's statement in Art. 62, ``That the surface of any elementary portion into which we may conceive the volume of the dielectric divided must be conceived to be charged so that the surface-density at any point of the surface is equal in magnitude to the displacement through that point of the surface reckoned inwards [into the dielectric]. If the displacement is in the positive direction, the surface of the element will be charged negatively on the positive side of the element, and positively on the negative side. These superficial charges will in general destroy one another when consecutive elements are considered, except where the dielectric has an internal charge, or at the surface of the dielectric.''   In Art. 111, Maxwell repeats, ``Conceive any portion of the dielectric, large or small, to be separated (in imagination)
from the rest by a closed surface, then we must suppose that on every elementary portion of this surface there is a charge measured by the total displacement of electricity through that element of surface reckoned inwards [into the dielectric].''   Maxwell is saying that if a portion of the dielectric is imagined isolated from any other polarization $\vD$, or if the dielectric is divided into volume elements separated by infinitesimally thin shells in which there is no displacement ($\vD=0$), that is, no Maxwell electric polarization, then with $\vD$ maintained at its original value in the volume elements there is negative surface charge ($-D_n$) on the positive side of the volume element and positive surface charge ($+D_n$) on the negative side of the volume element, where $\vnh$ denotes the positive direction; see Fig. \ref{figVolumeElements}.  Incidentally, this argument holds for a general $\vD$ such as  in polarized anisotropic material and, as mentioned above, Maxwell allowed the polarized media to have tensor permittivities, permeabilities, and conductivities.
\par
This interpretation is further confirmed at the beginning of Art. 325 where Maxwell summarizes, ``We have seen that when electromotive force [$\vE$] acts on a dielectric medium it produces in it a state which we have called electric polarization, and which we have described as consisting of electric displacement [$\vD$] within the medium in a direction which, in isotropic media, coincides with that, of the electromotive force, combined with a superficial charge on every element of volume into which we may suppose the dielectric divided, which is negative on the side towards which the force acts, and positive on the side from which it acts.''  In a modern-day version of Maxwell's separated dielectric volume elements, we would imagine infinitesimally thin free-space (vacuum) separation shells in which $\vP=0$ but not $\vD=0$ because, unlike Maxwell, we would not consider the vacuum/ether as a dielectric containing electric polarization.  In the case of volume elements separated by free space, $D_n$ is continuous across the dielectric-vacuum interface and $\sigma_e =0$ on the dielectric surfaces.
\par
Although Maxwell didn't specify how the displacement $\vD$ could be made zero in the thin-shell region surrounding the volume elements (see Fig. \ref{figVolumeElements}), one could imagine this being done by inserting perfectly electrically conducting (PEC) material into the thin-shell region (while keeping the $\vD$ fixed within the volume elements) and then letting the induced surface charge bleed onto the surfaces of the dielectric volume elements.  
\par
\begin{figure}[htbp]
  \centering
  \includegraphics[width=80mm]{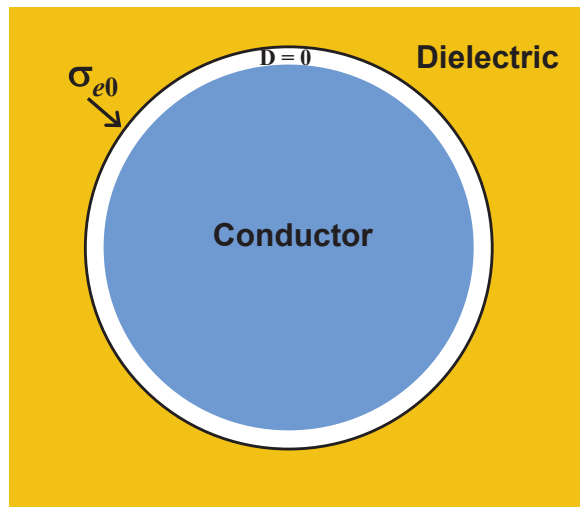}\\[-3mm]
  \caption{\small Charged conducting sphere separated from a dielectric material by an infinitesimally thin concentric shell in which $\vD = 0$ (zero Maxwell electric polarization).  The surface charge on the conducting sphere is canceled by the opposite charge on the inner surface of the concentric zero-$\vD$ shell, and the charge on the outer surface of the concentric zero-$\vD$ shell is transferred to the inner surface of the dielectric.}
  \label{figSphere}
\end{figure}

In discussing the Leyden jar, Maxwell explains how, according to his definition of electric polarization as displacement $\vD$, the charge on the surface of a conductor should be ultimately viewed as a surface charge on the surrounding dielectric (which can be a solid dielectric or the vacuum/ether): ``The charge therefore at the bounding surface of a conductor and the surrounding dielectric, which in the old theory was called the charge of the conductor, must be called in the theory of induction [$\vE$ inducing charge separation and electric polarization $\vD$]  the surface charge of the surrounding dielectric.''  Again, Maxwell is imagining an infinitesimally thin shell of zero electric polarization ($\vD=0$) separating the dielectric (which can be the vacuum/ether) from a conductor that, for example, can be a sphere with a surface charge density $\sigma_{e0}$.  Then the value of $\vD$ on either side of the external surface of the conductor will be zero and thus $\sigma_e$ becomes zero at the surface of the conductor.  However just inside the dielectric, the value of the radial displacement is equal to $\sigma_{e0}$ and thus the surface of the dielectric contains the surface charge $\sigma_{e0}$ because $\vD=0$ in the infinitesimally thin shell that borders the inner surface of the dielectric; see Fig. \ref{figSphere}.  Again, one could imagine $\vD$ being made zero in the thin shell by inserting PEC material into the thin shell and letting the induced surface charge on the inner surface of the PEC shell cancel the surface charge on the original conductor, then letting the surface charge on the outer surface of the PEC shell bleed onto the inner surface of the dielectric.
\section{Conclusion}
Although a source of consternation for past and present-day commentaries and histories of Maxwell's Treatise,  I would assert that none of Maxwell's explanations concerning surface charge are invalid given that his volume elements are separated not by infinitesimally thin free-space shells having zero $\vP$ and nonzero $\vD$, as one would  imagine today, but by infinitesimally thin shells having zero $\vD$.  As discussed above, Maxwell did this because he assumed that $\vD$ was the measure of electric polarization even in a free-space vacuum, Maxwell's ether.  Although it has been more than 150 years since Maxwell published his work in electromagnetics, as far as I am aware, this important interpretation of electric polarization as $\vD$ in Maxwell's work has not been previously pointed out or documented except very briefly in \cite{Yaghjian-Reflection}.
\ack
This research was supported in part under the U.S. Air Force Office of Scientific Research (AFOSR) Grant \# FA9550-19-1-0097 through Dr. Arje Nachman.
\end{document}